\documentclass{elsart5p}

\usepackage{graphicx}
\usepackage{amssymb,amsmath}

\newcommand {\mcu}{\mathcal{U}}

\newcommand{\cm}{\mathrm{c\!\:\!.m\!\:\!.}}
\newcommand{\He}{{}^3\mathrm{He}}

\newcommand{\Hh}{{}^3\mathrm{H}}
\newcommand{\nH}{n\text{-}{}^3\mathrm{H}}

\newcommand{\nHe}{n\text{-}{}^3\mathrm{He}}

\renewcommand{\vec}[1]{{\rightarrow{#1}}}

\begin{document}

\begin{frontmatter}
\title {Deuteron-deuteron scattering above four-nucleon breakup threshold}
 
\author[itpa]{A.~Deltuva},
\ead{arnoldas.deltuva@tfai.vu.lt}
\author[cfnul]{A.~C.~Fonseca} 
\address[itpa]
{Institute of Theoretical Physics and Astronomy, 
Vilnius University, A. Go\v{s}tauto 12, LT-01108 Vilnius, Lithuania}
\address[cfnul]{Centro de F\'{\i}sica Nuclear da Universidade de Lisboa, 
P-1649-003 Lisboa, Portugal }


\begin{abstract}
Deuteron-deuteron elastic scattering and transfer reactions in the energy 
regime above four-nucleon breakup threshold are described by solving 
exact four-particle equations  for transition operators.
Several realistic nuclear interaction models are used, including the one with 
effective many-nucleon forces generated by the explicit 
$\Delta$-isobar excitation; the Coulomb force between protons is taken 
into account as well. Differential cross sections, deuteron analyzing powers,
outgoing nucleon polarization, and deuteron-to-neutron polarization
transfer coefficients are calculated at 10 MeV deuteron energy.
Overall good agreement with the experimental data is found.
The importance of breakup channels is demonstrated.
\end{abstract}


\end{frontmatter}

\section{Introduction \label{sec:intro}}

 The pursuit for numerical solutions of the three- and four-nucleon
 scattering problems has been, since the early seventies of the last century,
 one of the
 most challenging endeavors in nuclear reaction theory, following the
 development of formal exact N-body equations using momentum or
 configuration space representations
\cite{yakubovsky:67,grassberger:67}.
 Progress advanced slowly at
 first and limited to the use of separable representations of
 subsystem operators, but with the advent of powerful computational
 tools, both in terms of algorithms (spline interpolation, Pad\'{e}
 summation, special integration meshes and weights, etc) and
 hardware advances, three-nucleon (3N) calculations with realistic 
nucleon-nucleon (NN) force
 models reached state-of-the-art status in the early 1990's
 due to the effort of a number of independent groups
\cite{koike:86a,chen:89a,friar:89a,cornelius:90a,kievsky:96a}.  Due to its
 higher dimensionality and multichannel complexity, the four-nucleon (4N)
 scattering problem took twenty years longer to reach the same status
 as the three-nucleon system except for the calculation of breakup
 amplitudes. These developments
 are mainly due to the works of the Pisa
\cite{viviani:01a,kievsky:08a,viviani:10a,viviani:13a},
 Grenoble-Strasbourg 
\cite{lazauskas:phd,lazauskas:04a,lazauskas:09a,lazauskas:12a},
and  Lisbon \cite{deltuva:07a,deltuva:07b,deltuva:12c,deltuva:14a}
groups. Because the first two groups use the coordinate-space
 representation, they were able to include, not only realistic two-body
 interactions, but also realistic three-body force
 models. Nevertheless they have had a major difficulty in calculating
 multichannel reactions and going beyond breakup threshold,
 particularly when the Coulomb interaction is included between
 protons. The Lisbon group uses the momentum space Alt, Grassberger and
 Sandhas (AGS) equations for transition operators 
\cite{grassberger:67} that can be solved
 for multichannel reactions both below and above breakup and with the
 Coulomb force included. The only stumbling block has been the
 inclusion of irreducible three-body forces. As alternative the nuclear 
force model with explicit excitation of a nucleon to a $\Delta$ isobar
was used. This coupling
 generates both effective three- and four-nucleon forces 
(3NF and 4NF) that have
 been successfully included in 4N calculations by the
 Lisbon-Hannover collaboration  \cite{deltuva:08a}.
The calculations using potentials derived from chiral effective field theory
have been performed as well \cite{deltuva:07a,deltuva:07b} but so far
including only the NN part of the interaction.

In the last 40 years progress in nuclear reaction theory has most
often succeeded experimental developments to the
point that, when calculations achieved a solid ground, the
instrumentation that gave rise to the data was no longer in
operation. Therefore inconsistencies between different data cannot
anymore be resolved by repeating the experiments or developing new
ones guided by the theoretical predictions. The 4N
scattering problem has suffered from this much more than the
3N system for the reasons mentioned above. Nevertheless,
new 4N scattering calculations  are worth pursuing
because they lead the way to the solution on complex multiparticle
scattering problems, not just in nuclear physics but also in cold atom
physics \cite{deltuva:10c}.  

In this work we present first results for 4N
reactions initiated by the collision of two deuterons ($d$) at energies
above four-particle breakup threshold. In this energy domain there are a few
shallow resonances \cite{tilley:92a}; 
therefore one does not expect the same problems as
encountered in $\nH$ and $\nHe$ near threshold.  
However, in two aspects the theoretical description of the $d+d$ scattering 
is even more interesting and challenging. First, since deuterons are
loosely bound and spatially large systems, the scattering of two deuterons looks like
the collision of two identical halo nuclei. The coupling to breakup channels 
in such a system is considerably stronger than in nucleon-trinucleon
scattering.  This makes the  $d+d$ reactions computationally more difficult
since open breakup channels lead to most complicated singularities in the
kernel of scattering equations and, furthermore, 
 a larger number of partial waves is needed. Second, the deuteron being a spin one
particle also provides the opportunity to calculate a number of
tensor observables, both in $d+d$ elastic as well as in
${}^2\mathrm{H}(d,p)\Hh$  and ${}^2\mathrm{H}(d,n)\He$
 transfer reactions for which there is experimental data. The
most abundant set of the experimental data for $d+d$ reactions exists at
deuteron energy $E_d = 10$ MeV where not only 
differential cross section and analyzing powers but also 
 deuteron-to-neutron polarization
transfer coefficients have been measured \cite{salzman:74}
 establishing the ${}^2\mathrm{H}(d,n)\He$ reaction as an
efficient source for polarized neutrons.

In Section \ref{sec:dyn} we explain the AGS equations we use and how 
to solve them. 
In Section \ref{sec:el} we show results for $d+d$ elastic scattering, while 
in Section \ref{sec:tr} results for ${}^2\mathrm{H}(d,p)\Hh$  and 
${}^2\mathrm{H}(d,n)\He$ transfer reactions are presented. 
Conclusions are drawn in Section \ref{sec:concl}.

\section{Deuteron-deuteron scattering equations \label{sec:dyn}}

As in our previous studies of 4N scattering
\cite{deltuva:07a,deltuva:07c,deltuva:14a}, we take advantage 
of the isospin symmetry and treat protons and neutrons as identical
fermions. This enables the symmetrization of the 
transition operators $\mcu_{\beta\alpha}$
thereby reducing the number of components that are 
distinct according to two-cluster partitions \cite{deltuva:07a}.
As usual, $\alpha =1$ labels the 3+1 partition (12,3)4, while
$\alpha =2$ stands for the 2+2 partition (12)(34). 
To ensure the required full antisymmetry of the four fermion system,
the employed basis states have to be antisymmetric 
under the exchange of two particles in subsystem (12),
and also in the subsystem (34) for the 2+2 partition.
The 4N transition operators $\mcu_{\beta 2}$ describing deuteron-deuteron scattering
obey the AGS  integral equations
\begin{subequations} \label{eq:U}
\begin{align}  
\mcu_{12}  = {}&  (G_0  t  G_0)^{-1}  
 - P_{34}  U_1 G_0  t G_0  \mcu_{12} + U_2 G_0  t G_0  \mcu_{22} , 
\label{eq:U12} \\  
\mcu_{22}  = {}& (1 - P_{34}) U_1 G_0  t  G_0  \mcu_{12} . \label{eq:U22}
\end{align}
\end{subequations}
Here $P_{34}$ is the permutation operator of particles 3 and 4,
\begin{equation}
t = v +vG_0 t
\end{equation}
 is two particle transition operator for the pair (12)
interacting via potential $v$, while
$U_1$ and $U_2$ are symmetrized transition operators for
3+1 and 2+2 subsystems, respectively \cite{deltuva:07a}.
The dependence of all transition operators
on the available energy $E$ arises through  the free 
four-particle resolvent
\begin{equation}
G_0 = (E+i\varepsilon-H_0)^{-1},
\end{equation}
with  $H_0$ being the free Hamiltonian. 
The finite imaginary part
$i\varepsilon$ is introduced in the complex energy method
to avoid singularities in the kernel of AGS equations, but
the limit $\varepsilon \to +0$ is needed for physical amplitudes,
that are given by the on-shell matrix elements of 
the transition operators  $\mcu_{\beta\alpha}$ \cite{deltuva:07a}.
The $\varepsilon \to +0$ limit is obtained by the analytic continuation of the finite
$\varepsilon$ results using the point method \cite{schlessinger:68}.
The analytic continuation, however, is only accurate when using sufficiently small
$\varepsilon$ values at which
the kernel of the AGS equations, although formally being nonsingular, 
still shows a quasisingular behavior \cite{deltuva:12c}. 
These quasisingularities reflect the presence of open
$p+\Hh$, $n+\He$,  $d+d$, $d+n+p$, and $n+n+p+p$ channels. 
Their treatment is taken over from Ref.~\cite{deltuva:12c}
where a special method for numerical integration
absorbing the quasisingular factor into the integration weights 
was developed. This way the 
quasisingularities can be integrated accurately without increasing significantly
the number of grid points.
In the present calculations we obtain well converged results
using $\varepsilon$ between 1.2 and 3.0 MeV in 0.3 MeV steps
and about 30 grid points for the discretization of each momentum variable.
Note that due to a larger weight of breakup channels a bit
more $\varepsilon$ values (about 6) are needed for a reliable $\varepsilon \to +0$ 
extrapolation as compared to previous calculations \cite{deltuva:12c,deltuva:13c}.

Although we explore the isospin symmetry, we also account for the isospin violation effects
due to the $pp$ Coulomb repulsion and the hadronic charge dependence (CD) of the nuclear force.
These effects cause the two-nucleon transition matrix $t$ to couple the states with different
total isospin in both 3N and 4N systems. In 3N or 4N total isospin basis, the 
two-nucleon transition matrix $t$ is given by linear combinations of  $pp$, $np$, and $nn$ 
transition operators as described in Ref.~\cite{deltuva:14b}.
For the $pp$ pair beside the nuclear force also the screened Coulomb potential is added, 
enabling rigorous inclusion  of the Coulomb interaction in the deuteron-deuteron scattering
via the method of screening and renormalization \cite{taylor:74a,alt:80a,deltuva:07c}.
We obtain fully converged results calculating the Coulomb-distorted 
short-range part of the amplitudes with the screening radius $R=13$ fm. 
The direct unscreened Coulomb amplitude is present only in the elastic scattering;
it is added after the renormalization of the short-range amplitude.
The direct  Coulomb amplitude causes the $d+d$ elastic differential cross section to 
diverge in the forward and backward direction but 
is absent for transfer reactions that are only distorted by Coulomb \cite{deltuva:07c}.
Other electromagnetic effects as the magnetic moment interaction are not explicitly
included in the calculations, however, their short-range part is implicitly included
in the employed NN potentials that are fitted to the NN data. They are not isolated
in the present work but,  based on previous studies \cite{deltuva:07b,kievsky:04a}
where they have been found to affect the vector analyzing powers only and to decrease
with increasing energy, 
 at 10 MeV one could expect only minor effects.

We solve the AGS equations in the momentum-space
partial-wave framework following the methodology developed
in Refs.~\cite{deltuva:07a,deltuva:12c}. In this framework the AGS equations constitute
a large system of coupled integral equations in three continuous Jacobi momentum variables. 
The integrals are discretized using Gaussian quadratures
with special (standard) weights for quasisingular (nonsingular) integrands,
leading to a huge system of linear algebraic equations. Beside the number of grid points
that can be kept moderate, 
the size of the resulting system depends also on the number of included angular
momentum states. With respect to the number of partial waves needed for achieving convergence,
 the $d+d$ reactions are more demanding than those initiated by
the $p+\Hh$ or $n+\He$ collisions already calculated in Refs.~\cite{deltuva:14a,deltuva:14b}.
In terms of the basis states defined in Refs.~\cite{deltuva:12a,deltuva:14b}, when solving
the AGS equations we include 4N partial waves with orbital angular momenta
 $l_x$ up to 6 and $l_y,l_z$ up to 7, total angular momenta  of the 2N subsystem $j_x,j_y$
up to  6, total angular momentum  of the 3N subsystem $J_y$
up to  $\frac{13}{2}$, and total 4N angular momentum 
$\mathcal{J}$ up to 7.
Once the AGS equations are solved, for the calculation of elastic observables
it is sufficient to include only the initial and final $d+d$ states 
with $l_z \leq 4$. In contrast, transfer reactions ${}^2\mathrm{H}(d,p)\Hh$
and ${}^2\mathrm{H}(d,n)\He$ require $l_z$ at least up to 6
in the channel states.

\section{Elastic scattering \label{sec:el}}

As in our previous calculations of nucleon-trinucleon scattering,
we use  several realistic nuclear force models, 
enabling us to study the sensitivity of the predictions to the dynamic input.
Beside two purely nucleonic interaction models,
 the inside-nonlocal outside-Yukawa
(INOY04) potential  by Doleschall \cite{doleschall:04a,lazauskas:04a} and
the CD Bonn potential \cite{machleidt:01a}, we also use 
the two-baryon potential CD Bonn + $\Delta$ \cite{deltuva:03c} that 
is the coupled-channel extension of CD Bonn, explicitly allowing  virtual
 excitation of a nucleon to a $\Delta$ isobar 
and thereby yielding mutually consistent effective three- and four-nucleon forces.
 This model, however, is not fitted to the trinucleon binding energy (BE),
yielding 7.53 (8.28) MeV for $\He$ ($\Hh$) which is increased relative to the CD Bonn BE 
result  of 7.26 (8.00) MeV. Only the  INOY04 model,
predicts the BE of $\He$ ($\Hh$) to be 7.73 (8.49) MeV, nearly reproducing the experimental
value of 7.72 (8.48) MeV. Since the $p+\Hh$ and $n+\He$
threshold positions depend on the respective BE, some scattering observables are expected to 
correlate with the BE, thereby establishing INOY04 as a reference potential.
Of course, the dependence of the observables on the used interaction model is in general 
much more complicated, but in particular cases simple correlations with 
nonmeasurable bound state properties such as the deuteron $D$-wave probability $P_D$ 
may take place. For the INOY04 potential $P_D=3.60$ \% while CD Bonn and CD Bonn + $\Delta$ 
have $P_D=4.85$ \%.

\begin{figure}[!]
\begin{center}
\includegraphics[scale=0.64]{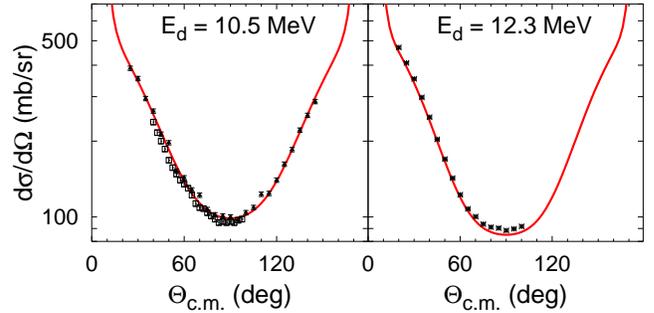}
\end{center}
\caption{\label{fig:els} 
Differential cross section of $d+d$ elastic scattering at 10.5 and 12.3 MeV
deuteron energy. Results calculated using INOY04 potential are compared 
with experimental data  from 
Refs.~\cite{wilson:69} ($\Box$), \cite{rosen:52} ($\blacktriangle$),
and \cite{PhysRevC.10.54} ($\times$).
}
\end{figure}

In the present work we concentrate on the $d+d$ scattering 
at deuteron energy $E_d = 10$ MeV where the 
most abundant set of experimental data exists, both for elastic
and transfer reactions.
The differential cross section $d\sigma/d\Omega$  for elastic  $d+d$ scattering
is, however, an exception. We therefore show $d\sigma/d\Omega$ 
as a function of the center of mass (c.m.) scattering angle $\Theta_{\cm}$
in Fig.~\ref{fig:els}  at $E_d = 10.5$ and 12.3 MeV.
Solely the INOY04 potential is used for predictions but, based on the study at $E_d =10$ MeV,
the sensitivity of this observable to the force model is small.
The angular dependence of $d\sigma/d\Omega$  is simple, with forward and backward peaks, 
where $d\sigma/d\Omega$ diverges due to the long-range Coulomb amplitude, and  
a minimum at $\Theta_{\cm}= 90^\circ$. This shape remains almost constant while the absolute
value of the differential cross section decreases with increasing energy.
Regarding the description of the experimental data, the 
picture is a bit contradictory. At $E_d = 10.5$ MeV there is a good agreement with the data from 
Ref.~\cite{rosen:52}  while the data from Ref.~\cite{wilson:69} are slightly
overpredicted, by 3\% at $\Theta_{\cm} = 90^\circ$.  
At $E_d = 12.3$ MeV the data from Ref.~\cite{PhysRevC.10.54} 
are well reproduced by the calculations at $\Theta_{\cm} < 65^\circ$ but slightly 
underpredicted around the minimum, by 4\% at $\Theta_{\cm} = 90^\circ$.  
These findings suggest that more calculations over a wider energy range
need to be performed and compared with the available data to determine the discrepancies 
between theory and experiment and  find out possible inconsistencies between data sets.

\begin{figure}[!]
\begin{center}
\includegraphics[scale=0.6]{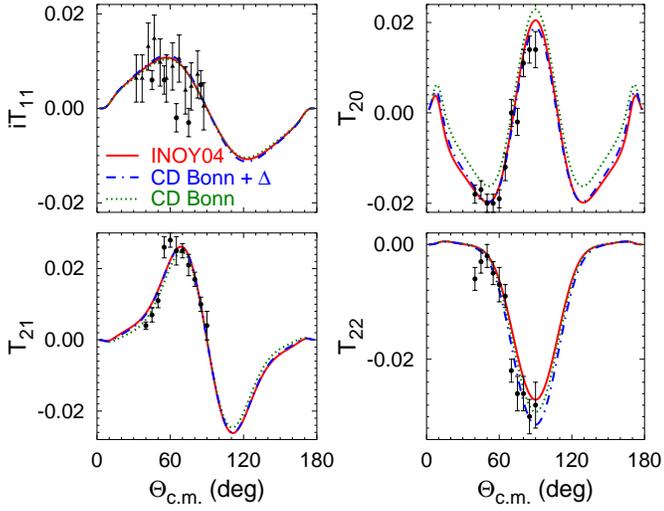}
\end{center}
\caption{\label{fig:elt} 
Deuteron analyzing powers in $d+d$ elastic scattering at $E_d = 10$ MeV.
  Results obtained with potentials INOY04
  (solid curves),  CD Bonn + $\Delta$ (dashed-dotted 
curves), and CD Bonn (dotted curves) are  compared with data  from 
Refs.~\cite{gruebler:72b} ($\bullet$) and \cite{plattner:69a} ($\blacktriangle$).
}
\end{figure}

In Fig.~\ref{fig:elt} we present results for deuteron
vector analyzing power $iT_{11}$ and tensor analyzing powers $T_{20}$, $T_{21}$, and $T_{22}$
in $d+d$ elastic scattering  at $E_d = 10$ MeV. 
The predictions are obtained using the potential models INOY04, 
CD Bonn + $\Delta$, and CD Bonn.  
These spin observables are very small 
in their absolute value, of the order of 0.02. 
Due to the identity of the two deuterons the angular distributions of elastic observables
in the c.m. frame are either symmetric ($d\sigma/d\Omega$, $T_{20}$, $T_{22}$)
or antisymmetric ($iT_{11}$, $T_{21}$) with respect to  $\Theta_{\cm}= 90^\circ$.
The overall description of the experimental data is good. The data have relatively large
error bars, especially in the $iT_{11}$ case where two data sets \cite{gruebler:72b,plattner:69a}
are available. The symmetric  analyzing powers $T_{20}$ and $T_{22}$ 
are most sensitive to the employed force model. However, the results obtained with different
potentials do not correlate with the properties of 2N and 3N bound states such as binding
energies or deuteron $D$-state probability.
The data are reproduced best by the 
predictions using the CD Bonn + $\Delta$ potential. 
 The $\Delta$-isobar effect is especially significant and beneficial for $T_{20}$.
The antisymmetric  analyzing powers $iT_{11}$ and $T_{21}$ show less sensitivity
to the employed potential. At least to some extent this is due to kinematic reasons,
since $iT_{11}$ are $T_{21}$ vanish exactly at $\Theta_{\cm}= 90^\circ$ where
$d\sigma/d\Omega$ has its minimum and the model dependence of symmetric  analyzing powers
reaches its maximum.

\section{Transfer reactions   \label{sec:tr}}

\begin{figure}[!]
\begin{center}
\includegraphics[scale=0.57]{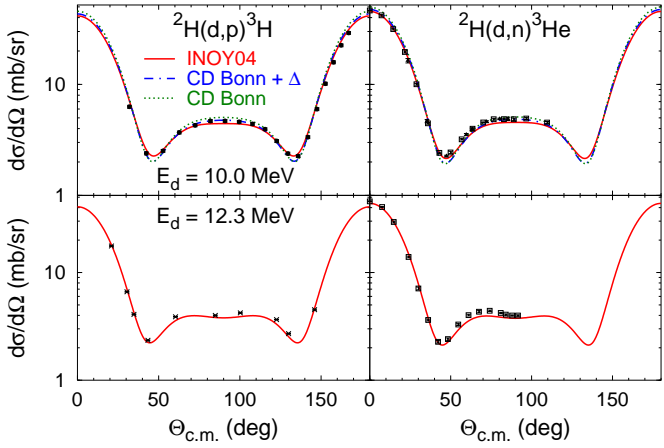}
\end{center}
\caption{\label{fig:trs} 
Differential cross section of ${}^2\mathrm{H}(d,p)\Hh$ (left) and ${}^2\mathrm{H}(d,n)\He$ (right)
transfer reactions at 10.0  and 12.3 MeV deuteron energy.
Curves as in Fig.~\ref{fig:elt}. The data are from 
Refs.~\cite{gruebler:72a} ($\bullet$), \cite{PhysRevC.10.54} ($\times$), 
\cite{drosg:78} ($\Box$), and \cite{thornton:69} ($\blacktriangle$).}
\end{figure}

We calculate the transfer reactions 
${}^2\mathrm{H}(d,p)\Hh$  and ${}^2\mathrm{H}(d,n)\He$ 
using the same nuclear interaction models, i.e.,  INOY04, 
CD Bonn + $\Delta$, and CD Bonn. 
In Fig.~\ref{fig:trs} we present  results at $E_d = 10.0$ and 12.3 MeV
for the differential cross section 
$d\sigma/d\Omega$ as function of the nucleon scattering angle $\Theta_{\cm}$
in the c.m. frame. 
For both ${}^2\mathrm{H}(d,p)\Hh$  and ${}^2\mathrm{H}(d,n)\He$ 
reactions,  $d\sigma/d\Omega$ has very similar shape but is slightly higher for the latter.
The differential cross section is symmetric with respect to $\Theta_{\cm}= 90^\circ$ but has a more
complicated angular and energy dependence than in the case of elastic scattering.
At both considered energies there are forward and backward peaks 
as well as local minima around $\Theta_{\cm}= 45^\circ$ and $135^\circ$.
At $E_d = 10.0$ MeV there is just a local maximum located at $\Theta_{\cm}= 90^\circ$ which
evolves into a shallow local minimum as the energy increases to
12.3 MeV; meanwhile two local maxima appear around $\Theta_{\cm}= 70^\circ$ and $110^\circ$.
 Such a behavior is seen also in the experimental data 
\cite{PhysRevC.10.54,gruebler:72a,drosg:78,thornton:69}.
The overall agreement between theoretical results and the data is good, except at
intermediate angles where $d\sigma/d\Omega$ is small and the data are slightly underpredicted
by the INOY04 results, roughly by 6\% at  $\Theta_{\cm}= 90^\circ$.
The sensitivity to the employed potential models is studied at $E_d = 10.0$ MeV and
is visible around the extrema of $d\sigma/d\Omega$.
While at forward and backward peaks and central maximum the predictions roughly
scale with the trinucleon BE as already observed in previous calculations 
below breakup threshold \cite{deltuva:07c,deltuva:10a}, at the minima CD Bonn and CD Bonn + $\Delta$
results are indistinguishable. This may indicate a possible sensitivity to two-nucleon
isospin singlet partial waves since there CD Bonn and CD Bonn + $\Delta$ potentials are identical.

\begin{table}[!]
\begin{tabular}{l*{3}{@{\quad}r}}
\hline
& $\sigma_p$  & $\sigma_n$ & $\sigma_b$  \\ \hline
CD Bonn  & 83.4 & 87.1 & 110  \\
CD Bonn + $\Delta$ & 79.8 & 83.9 & 113  \\
INOY04 & 77.6 & 81.3 & 112 \\
\hline
\end{tabular}
\caption{ \label{tab:tcs}
Predicted total cross sections for 
${}^2\mathrm{H}(d,p)\Hh$  and ${}^2\mathrm{H}(d,n)\He$ 
transfer reactions and for breakup, labeled $\sigma_p$, $\sigma_n$, and
$\sigma_b$, respectively, all in millibarns, at 
10 MeV deuteron energy.}
\end{table}

The resulting total cross sections $\sigma_p$ and $\sigma_n$ at $E_d =10$ MeV 
are collected in Table \ref{tab:tcs} for all three potentials.
The total transfer  cross sections scale quite well with the trinucleon BE. 
In addition we present also the total breakup cross section $\sigma_b$, 
including both three- and four-cluster channels. It is calculated
using the optical theorem with finite screening radius $R$ before subtraction 
and renormalization of the elastic scattering amplitude because 
transfer and breakup operators are short-ranged and the respective total cross sections 
are unchanged by renormalization phases \cite{deltuva:05d}.
 Already at $E_d =10$ MeV $\sigma_b$  
exceeds $\sigma_p$ and $\sigma_n$, indicating the importance of breakup in 
$d+d$ collisions. Note that in $n+\He$ scattering breakup becomes
the dominant inelastic channel only above the neutron laboratory energy of 
23 MeV which roughly corresponds to $E_d = 28$ MeV.

\begin{figure}[!]
\begin{center}
\includegraphics[scale=0.57]{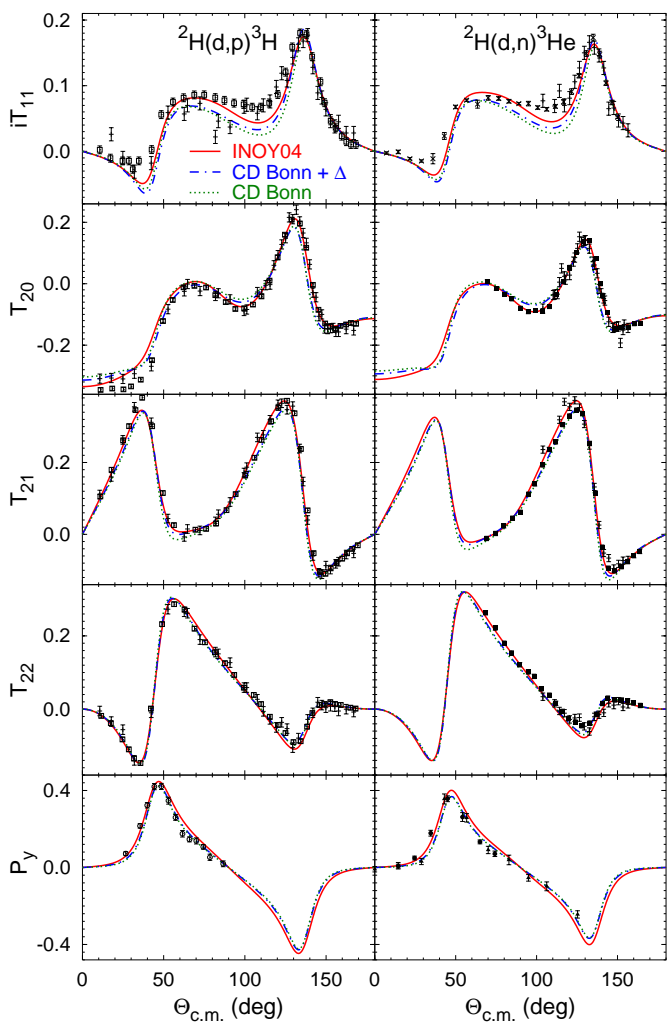}
\end{center}
\caption{\label{fig:trt} 
Deuteron analyzing powers and outgoing nucleon polarization 
of ${}^2\mathrm{H}(\vec{d},p)\Hh$ (left) and ${}^2\mathrm{H}(\vec{d},n)\He$ (right)
transfer reactions at 10 MeV deuteron energy.
Curves as in Fig.~\ref{fig:elt}. The data are from 
Refs.~\cite{gruebler:72a} (+), \cite{gruebler:81a} ($\Box$), \cite{guss:83} ($\times$), 
\cite{konig:79} ($\blacksquare$), \cite{hardekopf:72b} ($\circ$), \cite{spalek:72} ($\bullet$),
and \cite{salzman:74} ($\blacktriangle$).}
\end{figure}

Next we consider single-polarization spin observables in ${}^2\mathrm{H}(\vec{d},p)\Hh$
 and ${}^2\mathrm{H}(\vec{d},n)\He$ reactions at $E_d =10$ MeV. In  Fig.~\ref{fig:trt} 
we show vector analyzing power $iT_{11}$, tensor analyzing powers $T_{20}$, $T_{21}$, and $T_{22}$,
and outgoing nucleon polarization $P_y$ calculated using INOY04, 
CD Bonn + $\Delta$, and CD Bonn potentials.
All deuteron analyzing powers exhibit a complex angular dependence with
several local minima and  maxima. They show no symmetry with respect to 
$\Theta_{\cm}= 90^\circ$, in contrast to  $P_y$ which is antisymmetric.
The differences between ${}^2\mathrm{H}(\vec{d},p)\Hh$
 and ${}^2\mathrm{H}(\vec{d},n)\He$ observables are quite small but visible, e.g., around
second maximum of  $T_{20}$.  For most observables several measurements exist 
\cite{gruebler:72a,gruebler:81a,guss:83,konig:79,hardekopf:72b,spalek:72,salzman:74} 
that are at variance in particular cases,
especially for $iT_{11}$. Nevertheless, the overall description of the experimental data
by our calculations is successful with only few small or at most moderate disagreements,
mostly in the vector observables:
the minima of $iT_{11}$ are underpredicted while the positive peak of $P_y$ is shifted 
to larger angles.
The sensitivity to the used interaction model is quite small; it is most visible for
$iT_{11}$. It is not a simple scaling with trinucleon BE since often CD Bonn and CD Bonn + $\Delta$
predictions stay quite close together with INOY04 being further away.
This may again indicate the dominance of NN isospin singlet partial waves.
It may even indicate partial correlation of the observables with the deuteron $D$-state probability $P_D$, but more detailed studies are needed to confirm or reject this speculation.

\begin{figure}[!]
\begin{center}
\includegraphics[scale=0.6]{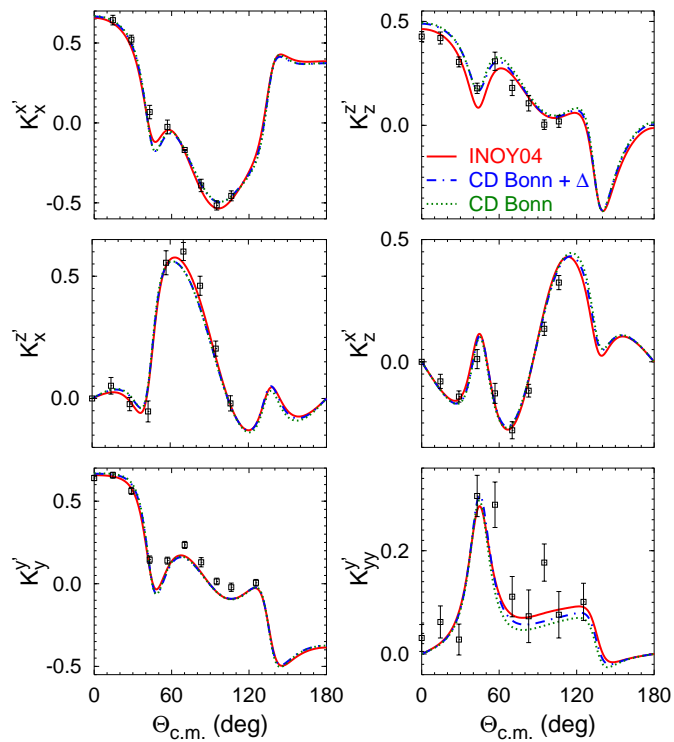}
\end{center}
\caption{\label{fig:trk} 
Deuteron-to-neutron polarization transfer coefficients 
of  ${}^2\mathrm{H}(\vec{d},\vec{n})\He$
reaction at 10 MeV deuteron energy.
Curves as in Fig.~\ref{fig:elt}. The data are from 
Ref.~\cite{salzman:74}.}
\end{figure}

Finally we show the results for double-polarization observables.  
Deuteron-to-neutron polarization transfer coefficients 
 $K_x^{x'}$, $K_x^{z'}$, $K_z^{x'}$, $K_z^{z'}$, $K_y^{y'}$, and $K_{yy}^{y'}$ 
in the ${}^2\mathrm{H}(\vec{d},\vec{n})\He$ reaction at $E_d = 10$ MeV 
have been measured in Ref.~\cite{salzman:74}.
This data and our predictions based on three nuclear interaction
models are compared in Fig.~\ref{fig:trk}. The polarization transfer coefficients
exhibit a very complex angular dependence having up to six local extrema.
Given such a complicated behavior of the observables the found agreement between theory
and experiment is impressive. There are only small discrepancies such as a slight underestimation
of $K_y^{y'}$ at intermediate angles.
The model dependence of the polarization transfer coefficients is quite weak
and qualitatively the same as already seen 
for single-polarization observables.

\section{Conclusions \label{sec:concl}}

We perform calculations for elastic and transfer reactions initiated by deuteron-deuteron
collisions above four-nucleon breakup threshold. This process mimics the scattering of two
halo nuclei. As dynamic input we use 
several realistic two-nucleon potentials and include the proton-proton Coulomb force 
via the screening and renormalization method.
Exact four-particle scattering equations in the integral form for symmetrized transition operators are
solved in the momentum-space framework where the presence of open breakup channels leads
to a kernel with a highly nontrivial  singularity structure.
The complex energy method with special integration weights is successfully applied to deal
with this difficulty. 
Compared to previous calculations of nucleon-trinucleon scattering, the relatively weak binding of 
deuteron and its large spatial size lead to additional complications such as a slower partial-wave
and complex-energy convergence.
Nevertheless, we obtain fully converged results for
$d+d$ elastic scattering as well as for 
${}^2\mathrm{H}(d,p)\Hh$  and ${}^2\mathrm{H}(d,n)\He$ transfer reactions.
For  these reactions at 10 MeV deuteron energy 
we calculate the differential cross section and all deuteron analyzing powers;
the former observable is predicted also at $E_d = 12.3$ MeV. Furthermore,
for transfer reactions we calculate also the outgoing nucleon polarization and,
in the ${}^2\mathrm{H}(d,n)\He$  case,
deuteron-to-neutron polarization transfer coefficients. 
The overall description of the experimental data is good, even for the most complicated
double-polarization observables. 
The comparison of predictions based INOY04, CD Bonn + $\Delta$, and CD Bonn potential models
may indicate  the dominance of NN isospin singlet partial waves
for most spin observables in transfer reactions, but not in the case of elastic scattering.
We also predicted the total breakup cross section and demonstrated the increased 
importance of breakup channels in $d+d$ reactions.

Together with the previous achievements in the nucleon-trinucleon scattering, 
the present work demonstrates that numerically exact calculations of all two-cluster reactions 
in the four-nucleon system
are now possible in a fully converged way using realistic nuclear interactions 
and including the pp Coulomb repulsion. 



\end{document}